\documentclass[10pt]{aa} 
\usepackage{graphics}
\begin{document} 
 
   \thesaurus{1          
	      (09.04.1;  
	       11.09.1 NGC 6946;
               11.19.2;  
               11.19.4;  
               11.19.6)} 

   \title{SCUBA observations of NGC 6946\thanks{
Based on observations at the James Clerk Maxwell
Telescope.  JCMT is operated by The
Joint Astronomy Centre on behalf of the Particle Physics and
Astronomy Research Council of the United Kingdom, the Netherlands
Organisation for Scientific Research, and the National Research
Council of Canada.
   }}
 
   \subtitle{}
 
   \author{
   S. Bianchi\inst{1,2}
   \and J. I.\,Davies\inst{1} \and P. B.\,Alton\inst{1} 
   \and M.\,Gerin\inst{3} \and F.\,Casoli\inst{4}} 
 
   \offprints{bianchi@mpia-hd.mpg.de}
 
   \institute{ Department of Physics \& Astronomy, Cardiff University,
   PO Box 913, Cardiff CF2 3YB, Wales, U.K. 
   \and Max-Planck-Institute f\"ur Astronomie, K\"onigstuhl 17, D-69117  
   Heidelberg, Germany.
   \and Radioastronomie millim\'etrique, ENS, 24 Rue Lhomond, F-75231
   Paris Cedex 05, France and UMR 8540 du CNRS
   \and DEMIRM, Observatoire de Paris, 61 Av. de l'Observatoire, F-75014
   Paris, France and UMR 8540 du CNRS
   } 
 
   \date{Received  / Accepted } 
 
   \maketitle 
 
   \begin{abstract} 

We present sub-millimetre images of the spiral galaxy NGC 6946 at 450 and 
850\,$\mu$m. The observations have been carried out using the scan/map
imaging mode of the Sub-millimetre Common User Bolometer Array (SCUBA) 
at the James Clark Maxwell Telescope (JCMT). The comparison with a
$^{12}$CO(2-1) line emission map from IRAM and with an U-band image,
clearly shows that the 850\,$\mu$m emission is associated with molecular gas
and bright star-forming regions, tracing the spiral arms. We place an
upper limit of 0.7 for the V-band optical depth of dust associated with
a diffuse atomic component.


\end{abstract} 

\section{Introduction}

Until recently,the main source of Far-Infrared (FIR) data for spiral
galaxies has come from the IRAS satellite.
The availability of instruments capable of observing at $\lambda>100\,\mu$m, 
like those onboard the satellite ISO, has allowed the detection of large 
amounts of cold dust, much colder than IRAS was able to detect. Spiral 
galaxies are found to have a dust content similar to the one in the Galaxy 
(Alton et al.~\cite{AltonA&A1998}).

Cold temperatures (T$<$20K) can be reached by diffuse dust heated by 
the general interstellar radiation field, while dust close to star-forming 
region is hotter (T$>$50K; Whittet \cite{WhittetBook1992}). 
Since diffuse dust is the main contributor
to the internal extinction in a galaxy, observations of cold dust help 
to trace its opacity. Additionally, ISO observations have
suggested that the dust distribution is more extended than the
stellar disk (Alton et al.~\cite{AltonA&A1998}; Bianchi, Davies \& Alton
\cite{BianchiPrep1999}). If this is confirmed, observation of the
distant universe may be severely biased, because of the large cross
section of the dust disks.

Unfortunately, the poor resolution ($\sim 2\arcmin$) of the current FIR
images does not allow detailed studies of the spatial distribution of
dust. Higher resolution can be achieved in the sub-mm, but a high
sensitivity is required because of the fainter dust emission. High
sensitivity and resolution are both characteristics of the recently
developed SCUBA sub-mm camera. Only a few large nearby galaxies have been
observed with SCUBA, notably the highly inclined galaxy NGC 7331 
(Bianchi et al. \cite{BianchiMNRAS1998}) and the edge-on galaxy
NGC 891 
(Alton et al. \cite{AltonApJL1998}; Israel et al. \cite{IsraelA&A1999}). 
The observed dust emission is found to correlate well with the molecular
gas phase, dominant in the centre. However, a dust component associated
with the atomic gas is needed to explain the dust and gas
column density at large galactocentric distance along the major axis of 
NCG~891 (Alton et al.~\cite{AltonSub1999}).

In this {\em Letter} we present SCUBA observations of the face-on
galaxy NGC~6946. Because the galaxy is larger than the camera field of
view, images have been produced with the scan-mapping technique,
chopping within the observed field. The observation and the data
reduction needed to restore the source signal are described in the next
section. The description and the discussion of the results are
given in Section 3. 

\section{Observations and data reduction}

NGC 6946 was observed at 450\,$\mu$m and 850\,$\mu$m, during April 10, 
11 and June 17, 18, 19, 20 1998.

SCUBA consists of two bolometer arrays of 91 elements optimised to
observe at 450\,$\mu$m and 37 elements optimised at 850\,$\mu$m, covering
a field of view of about 2.3 arcmin (Holland et al. 
\cite{HollandMNRASprep1998}). The camera, mounted on the Nasmyth focus
of the telescope can be used simultaneously at both wavelengths, by means 
of a dichroic beamsplitter.

In the scan-map mode, the telescope scans the source at a
rate of 24 arcsec per second, along specific angles to ensure a fully
sampled map. Meanwhile the secondary chops with a frequency of 7.8 Hz
within the observed field.
While this ensures a correct subtraction of the sky background,
the resulting maps unfortunately have the profile of the source
convolved with the chop. The profile of the source is restored deconvolving
the chop from the observed map by mean of Fourier Transform (FT) analysis.

Scan-maps of NGC~6946 presented here are fully sampled over an area of 
8$\arcmin$x8$\arcmin$. 
Each set of observations consisted of six scans, with different chop
configurations: chop throws of 20$\arcsec$, 30$\arcsec$ and 65$\arcsec$ 
along RA and Dec are needed to retrieve the final image.
Data have been reduced using the \textsc{STARLINK} package
\textsc{SURF}~(Jenness \& Lightfoot \cite{JennessMan1997}). 
Images were first flat-fielded to
correct for different sensitivities of the bolometers. Noisy bolometers
were masked and spikes from transient detections removed by applying a
5-$\sigma$ clip. A correction for atmospheric extinction was applied,
using measures of the atmosphere opacities taken several times during 
the nights of observation. Zenith optical depth varied during the six
nights, with $\tau_{450}=0.4-2.5$ and $\tau_{850}=0.1-0.5$. The 450\,$\mu$m
opacity on the last night was too high ($\tau>3$) for the source to be
detected and therefore the relative maps were not used for this
wavelength.
Because of the chopping in the source field, each bolometer sees a different 
background: a baseline, estimated from a linear interpolation at the edges 
of the scan, has been subtracted from each bolometer.

Sky fluctuations were derived from the time sequence of
observations for each bolometer, after the subtraction of a model of the
source, obtained from the data themselves. The images have then been
corrected by subtacting the systematic sky variations from each
bolometer.

Data taken with the same chop configuration were rebinned together into
a map in an equatorial coordinate frame, to increase the signal to
noise. Six maps with 3$\arcsec$ pixels were finally obtained for each
wavelength, combining 33 and 25 observations, at 850 and 450\,$\mu$m,
respectively.  In each of the six maps the signal from the source is 
convolved with a different chop function. The final deconvolved image is
retrieved using the Emerson II technique (Holland et al.  
\cite{HollandMNRASprep1998}; Jenness, Lightfoot \& Holland 
\cite{JennessProc1998}). Essentially, for each rebinned image, the 
FT of the source is derived by dividing the FT of the map by the FT of the
chop function, a simple sine-wave. Since the division boosts up the noise near
the zeros of the sine-wave, different chop configuration are used. 
For the chosen chop throws, the FT of the chop functions do not have
coincident zeros, apart from the zero frequency. A smoother FT of
source can therefore be obtained, and the final image is retrieved by
the applying an inverse FT.

Unfortunately the deconvolution introduces artifacts in the images,
like a curved sky background. This may be due to residual, uncorrected,
sky fluctuation at frequencies close to zero, where all the chops FT 
goes to zero. Work to solve this problem is ongoing
(Jenness, private communication). To enhance the contrast between the
sky and the source, we have modelled a curved surface from the images,
masking all the regions were the signal was evidently coming from the
galaxy. The surface has been then subtracted from the image.

Calibration was achieved from scan-maps of Uranus, that were reduced in
the same way as the galaxy. Integrated flux densities of Uranus were
derived, for each observing period, using the \textsc{STARLINK} package
\textsc{FLUXES} (Privett, Jenness \& Matthews \cite{PrivettMan1998})
for JCMT planetary fluxes.  Comparing data for each night we derived a
relative error in calibration of 8 per cent and 17 per cent, for 850\,$\mu$m 
and 450\,$\mu$m respectively. From the planet profile, the beam size was
estimated: FWHMs of 15.2$\arcsec$ and 8.7$\arcsec$ were measured for the beam at
850 and 450\,$\mu$m, respectively. To increase the signal to noise, the 
850\,$\mu$m image has been smoothed with a gaussian of 9$\arcsec$ thus
degrading the beam to a FWHM of 17.7$\arcsec$ The  450\,$\mu$m image has been
smoothed to the same resolution as for the 850\,$\mu$m one, to facilitate
the comparison between features present in both. The sky $\sigma$ in the 
smoothed images is 3.3 mJy beam$^{-1}$ at 850\,$\mu$m and 22 mJy beam$^{-1}$ 
at 450\,$\mu$m.

\begin{figure*}
\resizebox{\hsize}{!}{\includegraphics{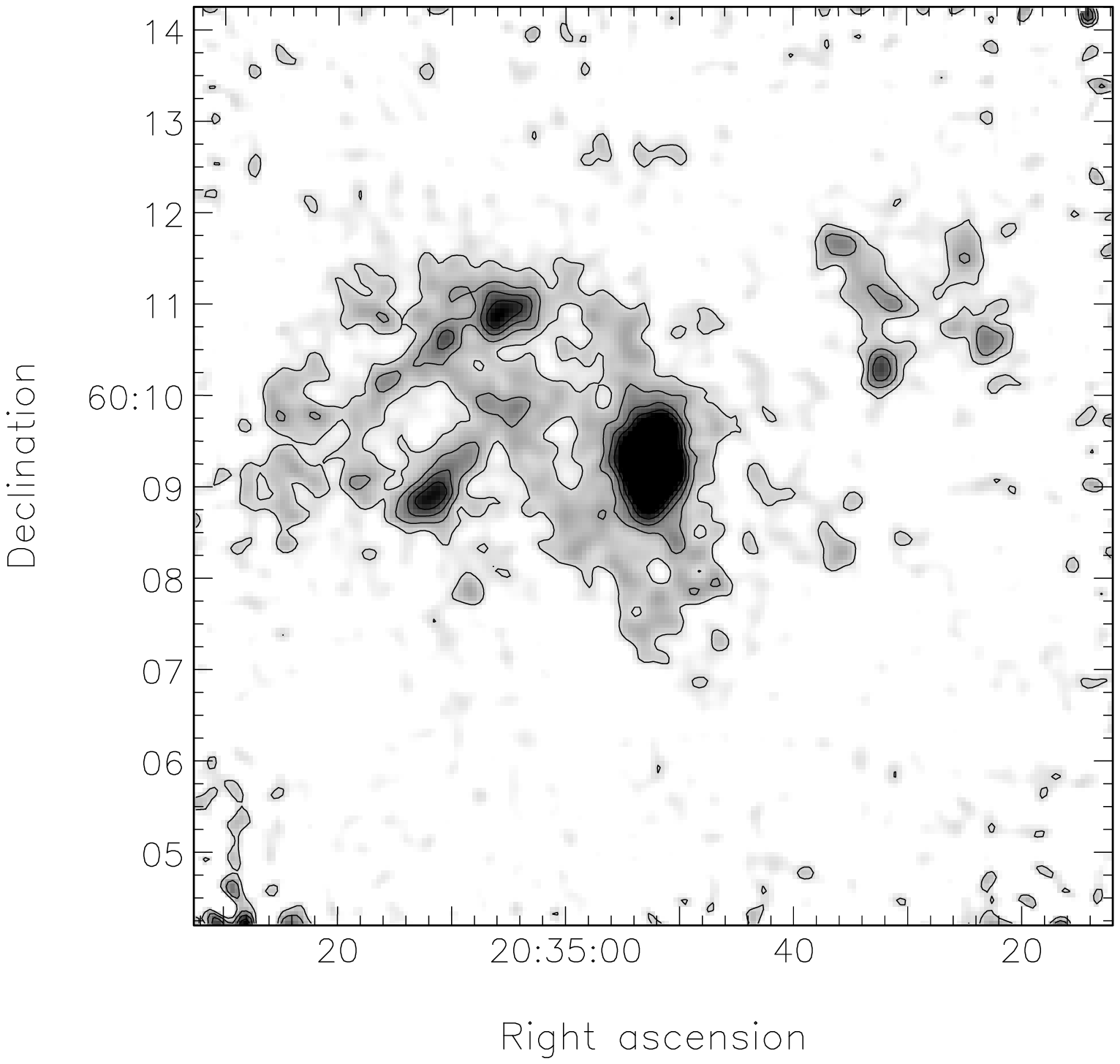}
		      \includegraphics{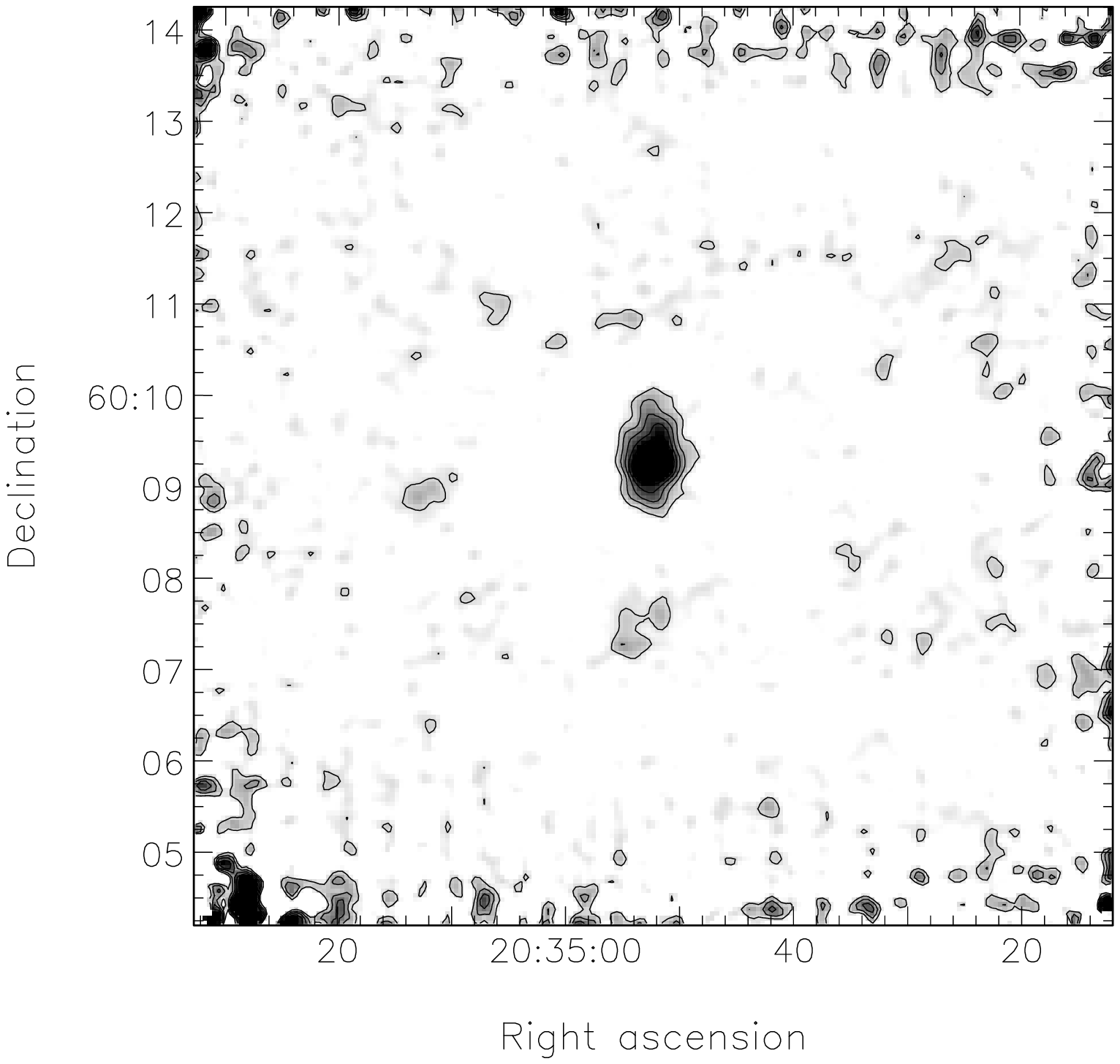}}
\resizebox{\hsize}{!}{\includegraphics{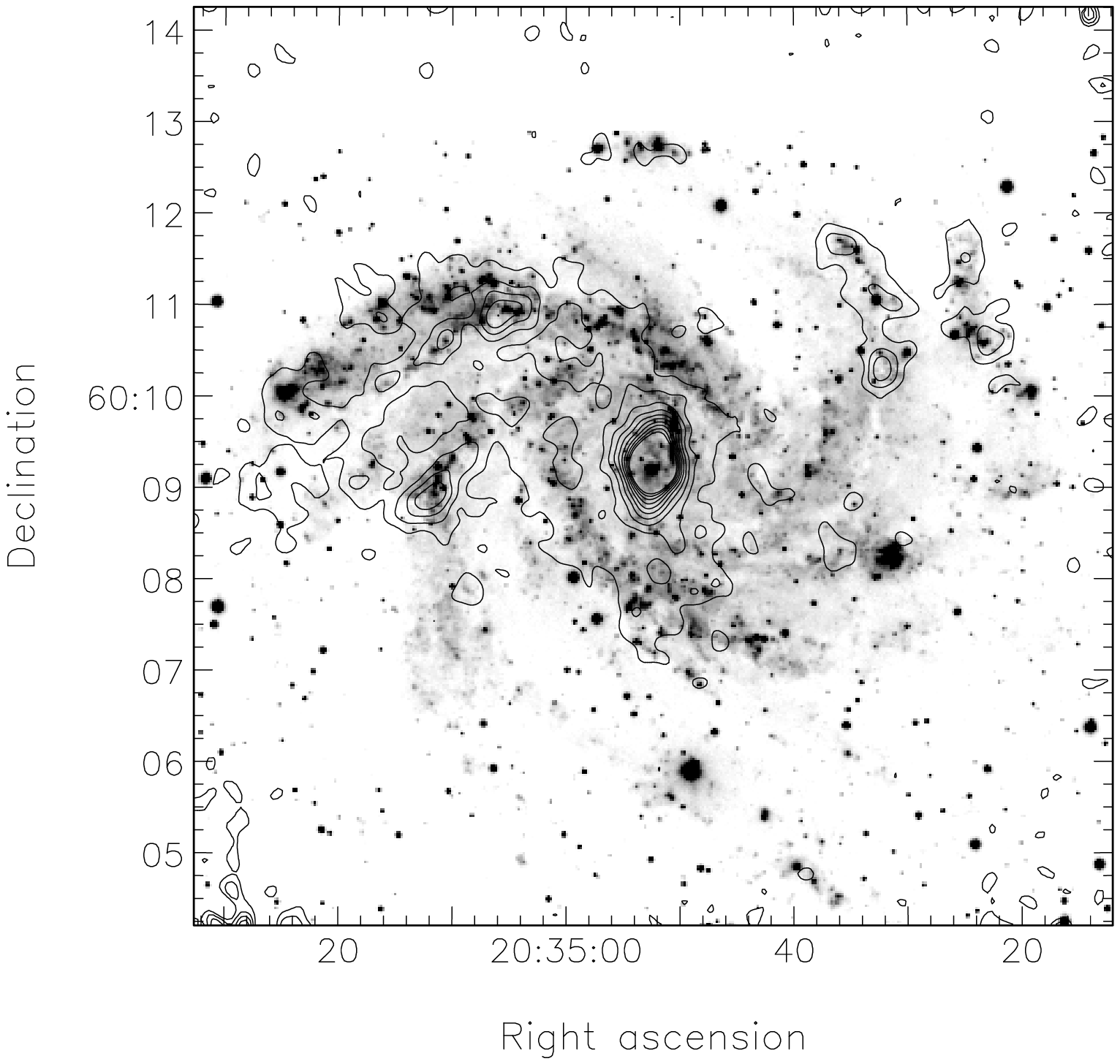}
		      \includegraphics{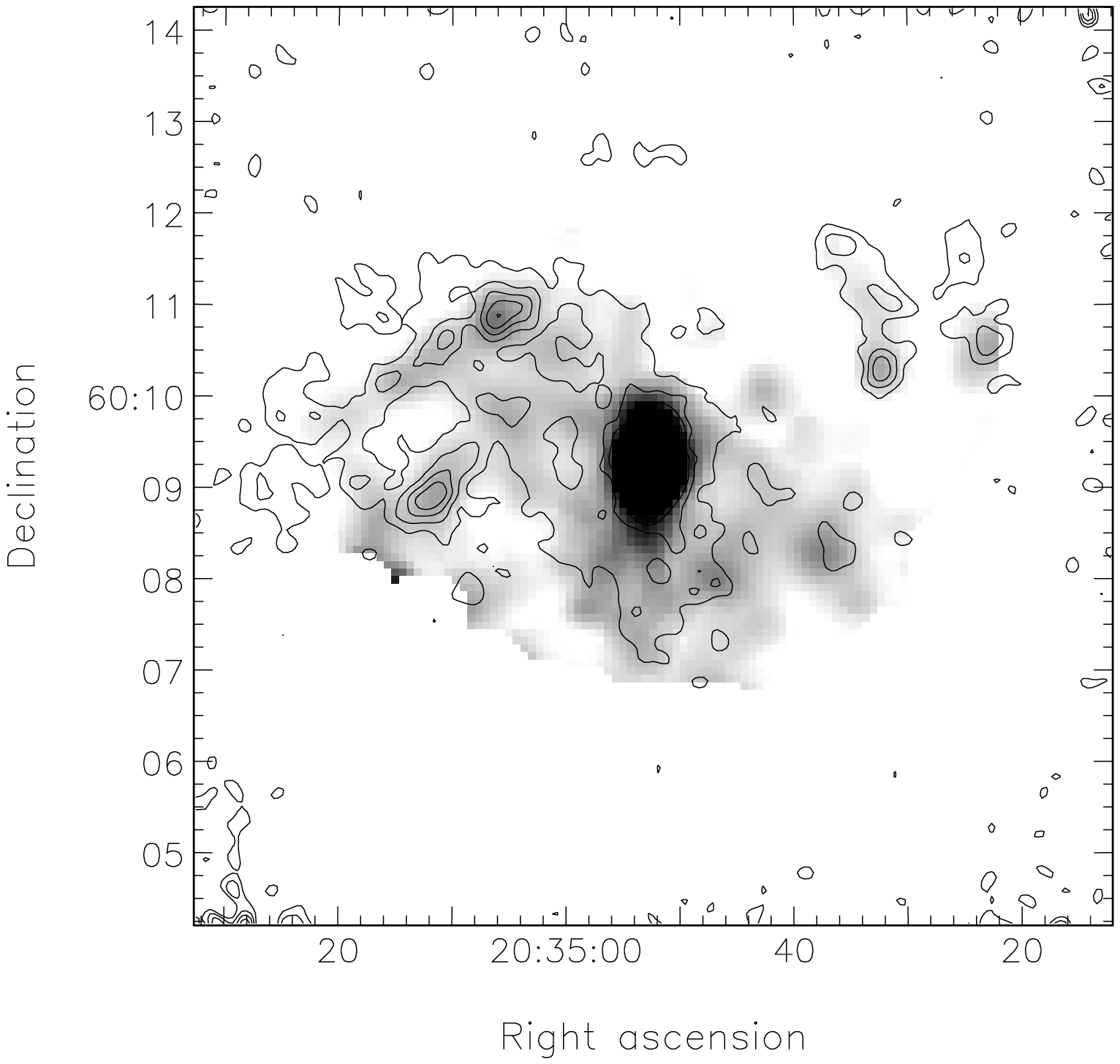}}
\caption{Sub-mm images of NGC 6946, at 850\,$\mu$m (top-left) and 450\,$\mu$m
(top-right). Grey scales show features 1-$\sigma$ above the sky, while
contours starts at 3-$\sigma$ and have steps of 3-$\sigma$. Both images
have a beam size FWHM=17.7$\arcsec$ An area of 10$\arcmin$x10$\arcmin$ is 
displayed, but only the central 8$\arcmin$x8$\arcmin$ are fully sampled. 
North is on top, East on the left.  
A U-band image of NGC 6946 (Trewhella \cite{TrewhellaThesis1998}) and a 
$^{12}$CO(2-1) emission map (Sauty, Gerin \& Casoli \cite{SautyA&A1998}),
are presented in the bottom-left and bottom-right panels, respectively.
850\,$\mu$m contours are overlayed to both optical and line emission
images.  The centre of the galaxy is in RA=20$^h$ 34$^m$ 52$\fs$3 and 
Dec=60$^\circ$ 09$\arcmin$ 14$\farcs$21 (J2000; De Vaucouleurs et al.
\cite{RC3}). The scale is 27 pc arcsec$^{-1}$ (D= 5.5 Mpc; Tully \cite{TullyBook1988}).}
\label{scuba_6946}
\end{figure*}

The final images, after removing the curved background and
smoothing are presented in Fig.~\ref{scuba_6946}. For each wavelength,
the grey scale shows all the features $>$1-$\sigma$, while contours starts
at 3-$\sigma$ and have steps of 3-$\sigma$.

\section{Results and discussion}

The 850\,$\mu$m image shows a bright nucleus and several features that 
clearly trace the spiral arms (in Fig.~\ref{scuba_6946} the sub-mm contours 
are overlayed to a U-band image of the galaxy 
(Trewhella \cite{TrewhellaThesis1998}))
As already seen in optical images (Tacconi \& Young 
\cite{TacconiApJ1990}), the spiral arms originating in the northeast 
quadrant are more pronounced than the others, where only regions with 
bright HII regions have detectable emission in the sub-mm. 
The 850\,$\mu$m image presents a striking similarity to the $^{12}$CO(2-1)
emission map in Sauty, Gerin \& Casoli (\cite{SautyA&A1998}), observed 
with the IRAM 30m radiotelescope with a comparable resolution (13$\arcsec$).
The image of molecular line emission is also shown in
Fig.~\ref{scuba_6946}, with the 850\,$\mu$m contour overlayed.
The similarity with the sub-mm image is hardly surprising, since the 
molecular gas is the dominant component of the ISM over the optical disk 
of NGC 6946 (Tacconi \& Young \cite{TacconiApJ1986}).
The nucleus is elongated in the direction north-south, as observed for
the central bar of molecular gas (Ishizuki et al. \cite{IshizukiApJ1990};
Regan \& Vogel \cite{ReganApJL1995}).

Emission associated with a more diffuse atomic gas component cannot be
detected, for several reasons. First of all the face-on inclination of
the galaxy: since dust is optically thin to its own emission a faint
component can be observed only if the dust column density is large.
This is the case for the high inclination galaxies NGC 7331 (Bianchi et 
al.~\cite{BianchiMNRAS1998})
and NGC 891 (Alton et al.~\cite{AltonApJL1998}), where higher signal to noise
were obtained coadding a smaller number of observations. The large
face-on galaxy M51 has been observed using the scan-map mode and
confirms the necessity of long integrations (Tilanus, private
communication).  Furthermore, 
chopping inside the source field removes not only the emission from the sky
but also from possible components with a shallow gradient: this may be the
case for dust associated with the flat HI distribution in NGC 6946 
(Tacconi \& Young \cite{TacconiApJ1986}). 
Finally, a faint diffuse emission could have been masked by the mentioned 
artifacts and subtracted together with the curved background.

The 450\,$\mu$m image is much noisier than the 850\,$\mu$m one,
because of the larger sky emission at this wavelength. Only a central
region of 0$\farcm$75 x 1$\farcm$5 can be clearly detected, although most 
of the features at a 3-$\sigma$ level correspond to regions emitting in 
the long wavelength image.

The temperature from the two sub-mm fluxes can be measured only for the
central region with significant 450\,$\mu$m flux. Sub-mm fluxes are 
1.2 Jy at 850\,$\mu$m and 9.3 Jy at 450\,$\mu$m. We checked for the
contribution of the strong $^{12}$CO(3-2) line emission at 346 GHz
to the 850\,$\mu$m flux using the observation of NGC~6946 centre
in this line reported by Mauersberger et al. (\cite{MauersbergerA&A1999}) 
for a beam of 21$\arcsec$. Converting from the original units to Jansky 
(Braine et al.~\cite{BraineA&A1995}) and averaging over the 30 GHz
bandwidth of 850\,$\mu$m filter (Matthews~\cite{MatthewsMan1999}), a
flux density of 80 mJy/beam is derived. As pointed out by the referee,
the pointing of the Mauersberger et al. observations was offset from the
strongly concentrated central emission by nearly one beamwidth in the SW
direction.  Using the $^{12}$CO(2-1) image as a template of the higher state 
emission, we corrected for the offset and derived the flux for the central 
region, larger than the beam.
A total contribution of 0.6 Jy is derived for the 346 GHz line (50\% of
the 850\,$\mu$m flux). However, this large contribution is due to the high 
density and gas temperature of the central region. 
In fact, for  850\,$\mu$m fluxes on larger apertures, the contamination is 
much smaller: Israel et al. (\cite{IsraelA&A1999}) derive a contribution of 
only 4\% to the total sub-mm flux of NGC~891. Therefore, the derivation of 
cold dust temperature at large galactic radii (Alton et 
al.~\cite{AltonApJL1998}) are not severely biased. We did not correct
the 450\,$\mu$m flux for the contribution of the $^{12}$CO(6-5) line,
that lies at the edge of the filter (Israel et al.~\cite{IsraelA&A1999}).

After the correction, the dust temperature of the central region is 
T=34$\pm$6 K, where the large quoted error comes from the calibration 
uncertainties. Here and in the following, dust temperature and masses are 
computed using the emissivity law $Q_{\mathrm{em}}(\lambda)$ derived by 
Bianchi, Davies \& Alton (\cite{BianchiA&A1999}) from observation of diffuse 
FIR emission and estimates of optical extinction in the Galaxy. 
For a wavelength dependence of the emissivity $\lambda^{-\beta}$, changing 
smoothly from $\beta=1$ to $\beta=2$ at 200\,$\mu$m (Reach et al. 
\cite{ReachApJ1995}) they obtain $Q_{\mathrm{ext}}(V)/Q_{\mathrm{em}}
(\mbox{100\,$\mu$m})=2390\pm190$, where $Q_{\mathrm{ext}}(V)$ is the
extinction efficiency in the V-band ($Q_{\mathrm{ext}}(V)\approx$1.5; 
Casey \cite{CaseyApJ1991}).  Lacking information outside of the centre, 
a mean temperature for a 
larger aperture can be derived from the lower resolution IRAS and ISO images at
100\,$\mu$m and 200\,$\mu$m (Alton et al.~\cite{AltonA&A1998}). The total
flux inside the B-band half light aperture (5$\arcmin$ in diameter) is 240$\pm$40 Jy
at 100\,$\mu$m and 280$\pm$40 Jy at 200\,$\mu$m (Bianchi, Davies \& Alton
\cite{BianchiPrep1999}). The temperature from IRAS and ISO fluxes is 
T=24$\pm$2 K.

\begin{figure}
\resizebox{\hsize}{!}{\includegraphics{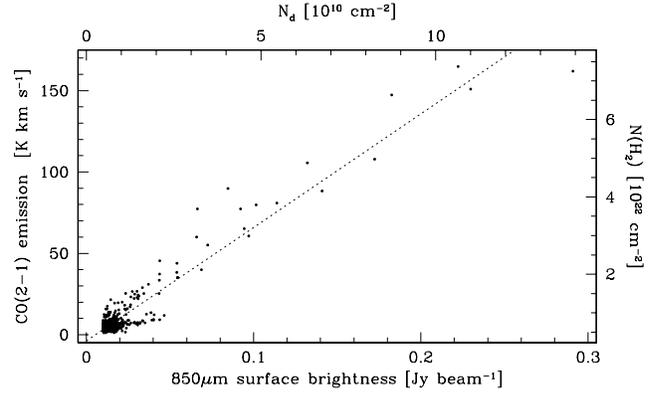}}
\caption{Point-to-point correlation between the $12$CO(2-1) line emission
and the SCUBA flux, for signals larger than 3-$\sigma$ in both
observations. Column density are derived as described in the paper.  }
\label{corre}
\end{figure}

We derived a point-to-point correlation between the  850\,$\mu$m flux
and the $^{12}$CO(2-1) line, resampling the sub-mm image to the
same pixel size as the line emission map (10$\arcsec$, roughly
equivalent to both beam sizes) and using all
positions with signals larger than 3-$\sigma$ in both observations.
A linear correlation is found (Fig.~\ref{corre}). 
Assuming a mean dust grain radius $a=$ 0.1$\mu$m and mass density
$\rho$=3 g cm$^{-3}$ (Hildebrand \cite{HildebrandQJRAS1983}),
the emissivity of Bianchi et al. (\cite{BianchiA&A1999})\footnote{
Using the Bianchi et al. emissivity for $Q_{\mathrm{em}}\propto\lambda^{-2}$ 
at any wavelength, results in dust column densities smaller by only 15\%.} 
and T=24K, the dust column density and 
hence the mass along the line of sight can be easily computed.
The molecular gas column density has been derived from the
$^{12}$CO(2-1) emission using a conversion factor appropriate for the
$^{12}$CO(1-0) emission in the general ISM in the Galaxy (X=1.8\,10$^{20}$
cm$^{-2}$ K$^{-1}$ km$^{-1}$ s; Maloney~\cite{MaloneyProc1990})
and a line ratio I(2-1)/I(1-0)=0.4 (Casoli et al.~\cite{CasoliA&A1990}).
The slope of the linear correlation can then be converted into
a gas-to-dust mass ratio of 170$\pm$20, a value very
close to the local Galactic one (160; Sodroski et al. \cite{SodroskiApJ1994}).
This confirms the association of dust with the local dominant phase of
the galactic ISM.

The dust content of NGC~6946 has been studied carrying out an energy
balance between the stellar emission in the optical and the FIR dust
emission, through the help of radiative transfer models. If an
exponential disk is used to model the dust distribution, a central 
face-on optical depth $\tau_V\sim5$ is needed to explain the FIR
emission (Evans~\cite{EvansThesis1992};
Trewhella~\cite{TrewhellaThesis1998}; Bianchi et
al.~\cite{BianchiPrep1999}). The 850\,$\mu$m image clearly show that
the dust distribution is more complex, but still the 
column densities derived from the sub-mm flux support the idea of an 
optically thick dust distribution. Under the same assumption of the 
previous paragraph, the diffuse component of the north-east spiral arms 
at a 3-$\sigma$ level corresponds to a V-band optical depth 
$\tau_V\approx 2.2$. The quite high optical
depth corresponding to the sky noise ($\tau_V\approx 0.7$) shows how
difficult is to obtain sub-mm images of dust emission in the 
outskirts of face-on galaxies, even for a high sensitivity 
instrument like SCUBA. Thus, possible extended dust distributions (Alton et
al.~\cite{AltonA&A1998}) are better revealed through deep sub-mm imaging
of edge-on galaxies, where the dust column density is maximized.
However, the high inclination makes the interpretation of the dust
emission along the line of sight more complex.

\begin{acknowledgements}
It is a pleasure to thank Gerald Moriarty-Schieven and Tim Jenness, for 
their support during observations and data reduction, respectively.
The paper has also benefited from the help of U. Klaas and R. Chini.
\end{acknowledgements}

\end{document}